%% file: main.tex
\documentclass[conference]{IEEEtran}
\usepackage[utf8]{inputenc}
\usepackage{cite}
\usepackage{clrs+} 
\usepackage{amsmath,amssymb,amsfonts}
\usepackage{algorithm, enumitem}
\usepackage[noend]{algpseudocode}
\usepackage{graphicx}
\usepackage{subcaption}
\usepackage{multirow}
\usepackage{booktabs}
\usepackage{colortbl}
\usepackage{hyperref}
\usepackage{flushend}
\usepackage[font=small,labelfont=bf]{caption}

\usepackage{xcolor}

\newcommand{\vect}[1]{\mathbf{#1}}
\algblock{ParFor}{EndParFor}
\algnewcommand\algorithmicparfor{\textbf{for}}
\algnewcommand\algorithmicpardo{\textbf{do in parallel}}
\algnewcommand\algorithmicendparfor{}
\algrenewtext{ParFor}[1]{\algorithmicparfor\ #1\ \algorithmicpardo}
\algrenewtext{EndParFor}{\textbf{end parallel for}}
\algnewcommand{\LineComment}[1]{\State \(\triangleright\) #1}
\newcommand{\mat}[1]{\mathbf{#1}}
\newcommand{\flop}{\mathrm{flops}}

\newcommand{\nnz}{\textit{nnz}}

\newcommand{\erdosrenyi}{Erd\H os-R\'{e}nyi}

\newcommand{\mA}{\mathbf{A}} 
\newcommand{\mB}{\mathbf{B}}

\newcommand{\transpose}     {^{\mbox{\scriptsize \sf T}}}

\newcommand{\mC}{\mathbf{C}}

\newcommand{\algoname}{\text{TS-SpGEMM~}}

\title{\huge{Distributed-Memory Parallel Algorithms for  Sparse Matrix and Sparse Tall-and-Skinny  Matrix Multiplication}}

\author{\IEEEauthorblockN{
    Isuru Ranawaka\IEEEauthorrefmark{1},
    Md Taufique Hussain\IEEEauthorrefmark{2},
    Charles Block\IEEEauthorrefmark{3},
    Gerasimos Gerogiannis\IEEEauthorrefmark{4},
    Josep Torrellas\IEEEauthorrefmark{5},
    Ariful Azad\IEEEauthorrefmark{6}}  
  \IEEEauthorblockA{\IEEEauthorrefmark{1}
    Indiana University, Bloomington, IN, USA (isjarana@iu.edu)}
    \IEEEauthorblockA{\IEEEauthorrefmark{2}
     Indiana University, Bloomington, IN, USA (mth@iu.edu)}
  \IEEEauthorblockA{\IEEEauthorrefmark{3}
    University of Illinois at Urbana-Champaign, IL, USA (coblock2@illinois.edu)}
    \IEEEauthorblockA{\IEEEauthorrefmark{4}
    University of Illinois at Urbana-Champaign, IL, USA (gg24@illinois.edu)}
    \IEEEauthorblockA{\IEEEauthorrefmark{5}
    University of Illinois at Urbana-Champaign, IL, USA (torrella@illinois.edu)}
    \IEEEauthorblockA{\IEEEauthorrefmark{6}
    Indiana University, Bloomington, IN, USA (azad@iu.edu)}
}  
\date{}

\begin{document}
\maketitle

\let\thefootnote\relax\footnotetext{Accepted as a conference paper at The International Conference for High Performance Computing, Networking, Storage, and Analysis 2024 (SC '24).}

\begin{abstract}
We consider a sparse matrix-matrix multiplication (SpGEMM) setting where one matrix is square and the other is tall and skinny. This special variant, 
{\em TS-SpGEMM}, has important applications in multi-source breadth-first search, influence maximization, sparse graph embedding, and algebraic multigrid solvers. 
Unfortunately,
popular distributed algorithms like sparse SUMMA deliver
suboptimal performance for TS-SpGEMM. To address this limitation, we 
develop a novel distributed-memory algorithm tailored for TS-SpGEMM. 
Our approach employs customized 1D partitioning for all matrices involved
and leverages sparsity-aware tiling for efficient data transfers. In 
addition, it minimizes communication overhead 
by incorporating both local and remote computations. 
On average, our  TS-SpGEMM algorithm attains 5$\times$ performance gains over 2D and 3D SUMMA. Furthermore, we use our algorithm to
implement multi-source breadth-first search and sparse graph embedding algorithms
and demonstrate their scalability up to 512 Nodes (or 65,536 cores) on NERSC Perlmutter.




\end{abstract}

\input{intro}

\input{methods}

\input{results}

\input{conclusion}

\bibliographystyle{IEEEtran}
\bibliography{Ref}

\end{document}

%% file: intro.tex
\section{Introduction}
Multiplication of two sparse matrices (SpGEMM) is a prevalent operation in scientific computing~\cite{bell2012exposing, devine2006parallel}, graph analytics~\cite{solomonik2017scaling, Azad2018,combblas2}, and machine learning~\cite{qin2020sigma}. 
Within these diverse applications, SpGEMM appears in three main variations: (a) $\mA\mA$, which involves squaring a sparse matrix $\mA\in \mathbb{R}^{n\times n}$ 
and is used in Markov clustering~\cite{Azad2018} and triangle counting~\cite{Azad2015}; 
(b) $\mA\mA\transpose$, which entails multiplying a matrix by its transpose  and
is useful in calculating Jaccard similarity~\cite{besta2019communication, hassani2023parallel}, performing sequence alignments~\cite{sc20pastis}, and conducting hypergraph partitioning~\cite{devine2006parallel}; and
 (c) $\mA\mB$, which involves multiplying two distinct sparse matrices  and 
 is used in multi-source breadth-first search (BFS)~\cite{then2014more}, the initial phase of Algebraic Multigrid solvers~\cite{bell2012exposing}, influence maximization~\cite{minutoli2019fast}, and generating sparse graph embeddings. 
In this paper, we focus on a special instance of the third variant referred to as Tall-and-Skinny-SpGEMM (TS-SpGEMM), where  $\mA\in \mathbb{R}^{n\times n}$ is a square matrix and $\mB\in \mathbb{R}^{n\times d}$ is a tall and skinny matrix with $d\ll n$.

TS-SpGEMM plays a crucial role in both graph analytics and scientific computing applications. For instance, iterations of multi-source BFS traversals on a graph are equivalent to TS-SpGEMM operations, where $\mA$ is the adjacency matrix of the graph and $\mB$ represents the BFS frontiers for all searches. Such multi-source BFS operations are central to calculations of influence maximization~\cite{minutoli2019fast} and closeness centrality~\cite{then2014more}, where the sparsity of $\mB$ may vary significantly across iterations. Additionally, TS-SpGEMM is applicable to sparse graph embedding algorithms, where each row of $\mB$ corresponds to a $d$-dimensional sparse embedding of a vertex, and is used in 
graph neural networks (GNNs) that support sparse embeddings~\cite{rahman2022triple}. In the context of Algebraic Multigrid (AMG) methods, TS-SpGEMM is utilized during the setup phase, where $\mB$ is the restriction matrix created from a distance-2 maximal independent set computation~\cite{bell2012exposing}.

Despite its wide range of applications, TS-SpGEMM has not received focused attention in the literature (see Table~\ref{table:experimental-setting}). Current distributed-memory SpGEMM algorithms, such as Sparse SUMMA~\cite{Buluc2012, azad2016exploiting} in CombBLAS ~\cite{combblas2} and 1-D partitioning-based algorithms in Trilinos~\cite{Heroux2005} and PETSc~\cite{Petsc2014}, perform well for standard scenarios but fall short in the TS-SpGEMM context, as demonstrated by our experiments. 

Our work aims to address this gap by introducing a scalable distributed-memory algorithm specifically designed for TS-SpGEMM.
Our algorithm follows the principles of Gustavson's algorithm~\cite{Gustavson1978}, which constructs the output row-by-row. 
Since both the $\mB$ matrix and the output matrix (which we
call $\mC$) are tall and skinny, 1-D partitioning is better suited for them.
For instance, in many cases, the number of columns in matrix $\mB$ 
is lower than the number of processes. 
However, a basic implementation of Gustavson's algorithm in distributed memory might necessitate fetching a substantial portion of $\mB$ into a process, potentially exceeding local memory capacity.
We mitigate this issue by utilizing a 2-D virtual layout for matrix $\mA$ and conducting multiplications tile by tile, where each tile represents a submatrix of $\mA$ stored within a process. During the execution of a tile, only the memory footprint of $\mB$ required by the tile needs to be stored in the local memory, reducing the concurrent memory footprint. By adjusting tile widths and heights, we can manage the granularity of computation and communication. 
 

We develop TS-SpGEMM with two computation modes. 
In the {\em local compute} mode, $\mA$ and  $\mC$  remain stationary, while data from $\mB$ is communicated to perform local multiplications. 
Conversely, in the {\em remote compute} mode, $\mB$ and $\mC$ remain stationary, while data from $\mA$  is transferred to remote processes where multiplication is performed, and partial results are then returned back to their respective processes.
The choice of local or remote computations is determined for each tile based on its sparsity pattern.
By balancing local and remote computations, we can reduce communication costs for certain applications. 
As for local computations, we 
adaptively select between 
a sparse accumulator (SPA)~\cite{gilbert1992sparse} or 
a hash-based accumulator~\cite{nagasaka2019performance} 
for local SpGEMM and merging of partial results.
These optimizations substantially enhance the performance of our TS-SpGEMM algorithm compared to existing distributed SpGEMM methods in CombBLAS and PETSc.

We have implemented two graph algorithms using TS-SpGEMM: multi-source BFS and sparse graph embedding. 
In multi-source BFS, the $\mC$ matrix from one iteration serves as the $\mB$ matrix in the subsequent iteration. Consequently, for scale-free graphs, the sparsity of $\mB$ fluctuates dramatically across BFS iterations. For sparse embedding, we implemented a force-directed graph embedding algorithm~\cite{Force2Vec}, maintaining the embedding sparse without compromising its quality. In both scenarios, TS-SpGEMM offered substantial performance benefits over other SpGEMM alternatives.

We summarize key contributions of this paper as follows:
\begin{itemize}
\item {\bf Algorithm}: We develop a distributed-memory algorithm for TS-SpGEMM. Our algorithm uses tiling to reduce memory requirements and
selectively employs local or remote computations to reduce communication. 

\item {\bf Comparison with SpMM and SUMMA}: We demonstrate the conditions under which TS-SpGEMM outperforms sparse and tall-and-skinny dense matrix multiplication (SpMM) and Sparse SUMMA.

\item {\bf Performance and Scalability}: For TS-SpGEMM (with $d=128$), our algorithm runs on average $5\times$  faster than alternative SpGEMM implementations in CombBLAS and PETSc. TS-SpGEMM scales well to 512 nodes (65,536 cores) on the Perlmutter supercomputer

\item {\bf Applications}: Multi-source BFS utilizing our TS-SpGEMM runs upto $10\times$ faster than SUMMA-enabled algorithms in CombBLAS.

\end{itemize}

\noindent
{\bf Availability}: TS-SpGEMM is publicly available at \href{https://github.com/HipGraph/DistGraph?tab=readme-ov-file#run-spgemm}{https://github.com/HipGraph/DistGraph}. The repository includes scripts to reproduce results presented in this paper.


%% file: methods.tex
\section{Background and Related Work}
\subsection{The TS-SpGEMM problem}\label{sec:ts-spgemm-prob}
The generalized sparse matrix multiplication (SpGEMM) multiplies two sparse matrices $\mA$ and $\mB$ and computes another potentially sparse matrix $\mC$.
In this paper, we consider TS-SpGEMM that multiplies a square matrix $\mA {\in} \mathbb{R}^{n \times n}$ with a tall and skinny matrix ${\mB {\in} \mathbb{R}^{n \times d}}$ and computes another tall and skinny matrix ${\mC {\in} \mathbb{R}^{n \times d}}$, where $d {\ll} n$. 
Although our TS-SpGEMM algorithms can cover the broader SpGEMM scenarios, our emphasis lies specifically on the TS-SpGEMM variant, which is used in numerous applications mentioned earlier.
TS-SpGEMM can be also performed on an arbitrary semiring $\mathbb{S}$ instead of the usual (x,+) semiring.
For example, we used a ($\land, \lor$) semiring in our implementation of multi-source BFS.
Given a matrix $\mA$, $\nnz(\mA)$ denotes the number of nonzeros in $\mA$, and 
$\flop$ denotes the number of multiplications needed to compute $\mA\mB$. 


\subsection{Related work}
SpGEMM is a well-studied problem with many sequential, shared-memory and distributed-memory parallel algorithms discussed in the literature. 
Gao et al.~\cite{gao2023systematic} provided an excellent survey of the field.


\textbf{Shared-memory parallel algorithms.}
Most shared-memory parallel SpGEMM algorithms can be categorized into two main classes. The first and widely adopted approach is based on Gustavson's algorithm~\cite{Gustavson1978}, which constructs the output column-by-column (or row-by-row)~\cite{nagasaka2019performance, patwary2015parallel, deveci2017performance, azad2016exploiting, davis2019algorithm}.
For performance, these algorithms rely on various accumulators based on heap~\cite{azad2016exploiting}, hash table~\cite{nagasaka2019performance}, and a dense vector called SPA~\cite{patwary2015parallel, gilbert1992sparse}.
The second approach uses the expand-sort-compress strategy~\cite{buluc2008representation, OuterSPACE-8327050, dalton2015toms, liu2019register, gu2020bandwidth}, which generates intermediate results through outer products of input matrices and then merges duplicated entries to obtain the final results. 
The performance of these algorithms depends on factors such as the sparsity of input matrices, compression ratio (the ratio of floating-point operations to nonzeros in the output matrix), number of threads, and efficient utilization of memory and cache.
Even in shared memory, most prior work evaluated algorithms with $\mA\mA$ and $\mA\mA\transpose$ settings.

\textbf{Distributed-memory parallel algorithms and libraries.}
Distributed-memory algorithms for SpGEMM can be classified based on the data distribution method they employ. 
Algorithms utilizing 1D partitioning distribute matrices across either the row or column dimension. 
Bulu\c{c} and Gilbert\cite{Buluc2008icpp} showed for the $\mA\mA$ case that 1D algorithm fails to scale due to communication cost.
The variant of 1D algorithm they considered forms output row-by-row while cyclicly shifting $\mB$ to avoid high memory cost.
To mitigate communication costs in 1D algorithms, preprocessing with graph/hypergraph partitioning models has been proposed~\cite{akbudak2018partitioning}. However, this preprocessing step can pose new scalability challenges, as it often does not scale efficiently.
In 2D distribution, the matrices are partitioned into rectangular blocks within a 2D process grid.
For example, CombBLAS~\cite{combblas2011} used  the Sparse SUMMA algorithm~\cite{Buluc2012, Geijn1995}, while Borvstnik et al.~\cite{borvstnik2014sparse} used Cannon's algorithm~\cite{cannon69thesis} 
on the 2D distribution of matrices. 
In the 3D (or 2.5D) variant of the Sparse SUMMA algorithm, each sub-matrix is further divided into layers.
This approach exhibits better scalability at larger node counts~\cite{azad2016exploiting, lazzaro2017increasing, batchedSpGEMMIPDPS21}, where the multiplied instances become more likely to be latency-bound.

Given the broad range of applications, most libraries covering sparse linear algebra include an implementation of SpGEMM. As illustrated in Table~\ref{table:library}, well-known libraries like CombBLAS~\cite{combblas2}, DBCSR~\cite{borvstnik2014sparse}, PETSc~\cite{Petsc2014}, Trilinos~\cite{Heroux2005}, and CTF~\cite{solomonik2015sparse} all feature various SpGEMM algorithms. 

\begin{table}[!t]
\centering
\caption{Libraries with implementations of distributed SpGEMM.}
\label{table:library}
\begin{tabular}{l l l  l l } 
\toprule
Library & Data Distribution & Algorithm \\
\toprule
CombBLAS~\cite{combblas2} & 2D, 3D & Sparse SUMMA  \\
DBCSR~\cite{borvstnik2014sparse} & 2D, 3D & Sparse Cannon, One sided MPI \\
Saena~\cite{rasouli2021compressed} & 1D & Recursive \\
PETSc~\cite{Petsc2014} & 1D & Distributed Gustavson \\
Trillinos~\cite{Heroux2005, nusbaum2011optimizing} & 1D & Distributed Gustavson \\
CTF~\cite{solomonik2015sparse} & 1D, 2D, 3D & Sparse SUMMA \\
\bottomrule
\end{tabular}
\end{table}

\begin{table}[!t]
\centering
\caption{Distributed SpGEMM algorithms with published experimental settings.}
\label{table:experimental-setting}
\begin{tabular}{l l l  l  l} 
\toprule
Algorithm & Data Distribution & Experiments \\
\toprule
Sparse SUMMA~\cite{Buluc2012, azad2016exploiting, combblas2} & 2D and 3D  & $\mA\mA$, $\mA\mA\transpose$, $\mA\mB$ \\
Sparse Cannon~\cite{borvstnik2014sparse} & 2D  & $\mA\mA$\\
Recursive~\cite{rasouli2021compressed} & 1D  & $\mA\mA$ \\
Hypergraph Partitioning~\cite{akbudak2018partitioning} & 1D & $\mA\mA$, $\mA\mA\transpose$, $\mA\mB$ \\
Survey~\cite{gao2023systematic} & 1D  & $\mA\mA$, $\mA\mA\transpose$ \\
TS-SpGEMM (this paper) & 1D (virtual 2D)  & $\mA\mB$ \\
\bottomrule
\end{tabular}
\end{table}

\textbf{Experimental settings.}
Over the years, researchers have evaluated distributed SpGEMM algorithms under various configurations of input matrices.
Table~\ref{table:experimental-setting} demonstrates that the majority of algorithms were evaluated for  $\mA\mA$ and $\mA\mA\transpose$ scenarios. 
Among these evaluations, 3D Sparse SUMMA~\cite{azad2016exploiting} and Akbudak et al.~\cite{akbudak2018partitioning} conducted experiments for $\mA\mB$ cases, where $\mB$ comprised rectangular matrices.
However, their experimental settings focused on the setup phase of the AMG solvers, where the number of columns in $\mB$ was typically substantial and frequently comparable to that of $\mA$.
Consequently, most existing algorithms have not been evaluated for TS-SpGEMM scenarios as discussed in this paper.

\section{Distributed-Memory Algorithms}
\begin{table}[!t]
\centering
\caption{List of notations used in the paper}
\begin{tabular}{l l} 
\toprule
\textbf{Symbol}                 & \textbf{Description}                         \\ 
\toprule
$\vect{A}$                      & The $n{\times}n$ square matrix, row-wise 1D partitioned  \\
$\vect{A}^c$                      & The $n{\times}n$ square matrix, column-wise 1D partitioned  \\
$\vect{B}$                      & The $n{\times} d$ tall and skinny matrix, row-wise 1D partitioned   \\
$\vect{C}$                      & The $n{\times} d$ output matrix, , row-wise 1D partitioned  \\
$P_{i}$                         & The $i$th process \\
$\mA_i$ & The $\frac{n}{p}{\times}n$ submatrix of $\mA$ stored at $P_i$\\

$\mA_i^c$ & The $n{\times}\frac{n}{p}$ submatrix of $\mA^c$ stored at $P_i$\\
$p$                             & The number of processes \\

$h$                             & The height of a computing tile \\

$w$  & The width of the  tile\\



$t$            & The number of threads\\

\bottomrule
\end{tabular}

\label{tab:notations}
\end{table} 

\begin{figure}
		\centering
		\includegraphics[width=0.99\linewidth]{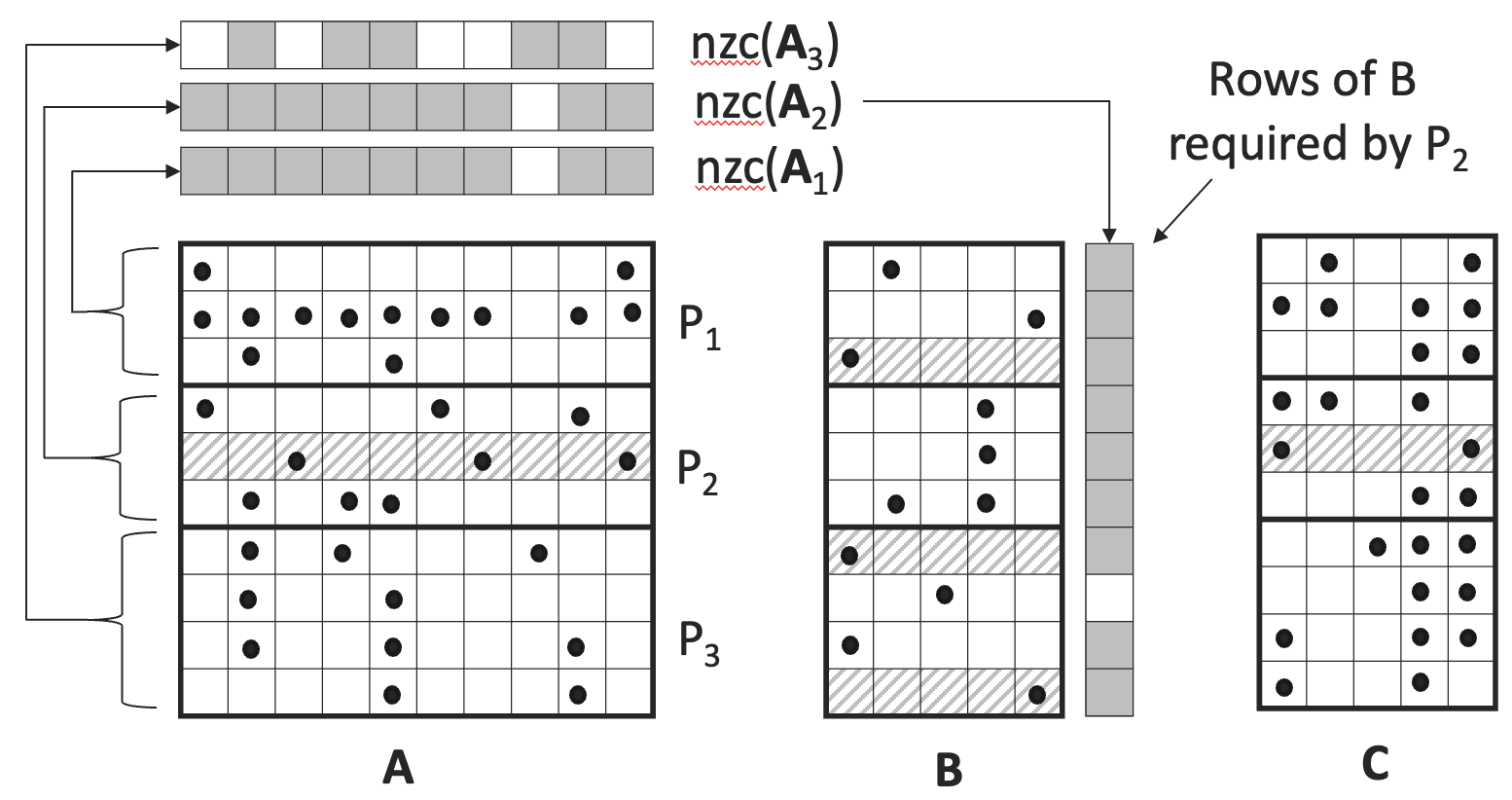}
        \caption{ 
        Distributed memory Gustavson's algorithm using a toy example of $10\times10$ square sparse matrix and $10\times5$ tall-and-skinny sparse matrix distributed over $3$ processes in a row partitioned manner (dark lines represent process boundary).
        Shaded rows of $\mA$ and $\mB$ represent parts of the input matrices involved in the computation of the shaded row of $\mC$.
        Shaded elements in the $nzc$ vectors represent columns with at least one non-zero (non-zero columns) of the local matrix owned by each process.
        Thus, the nzc vector of each process determines which rows of $\mB$ are accessed by this process.
        Note that while the sparsity patterns of $\mA_1$ and $\mA_2$ differ significantly, both require all but one row of $\mB$.
        }
		\label{fig:dist-gustavson}
\end{figure}

\subsection{Distributed TS-SpGEMM based on Gustavson's algorithm}
In the row variant of Gustavson's algorithm, to compute the $r$th row of $\mC$, the $r$th row of $\mA$ is scanned.
For each nonzero element stored at location $(r,c)$ of $\mA$, the $c$th row of $\mB$ is scaled with that particular non-zero value of $\mA$ and merged together to obtain $\mC(r,:)$ as follows:
\begin{equation}
\mC(r,:) = \sum_{c:\mA(r,c) \neq 0} \mA(r,c) \mB(c,:)
\label{eq:row-by-row}    
\end{equation}

\begin{algorithm}[!t]
\caption{Overview of Naive TS-SpGEMM}
\label{alg:ts-spgemm-naive}
\textbf{Input:} $\mathbf{A} \in \mathbb{R}^{n \times n}$ and $\mathbf{B} \in \mathbb{R}^{n \times d}$ distributed in $p$ processes.
\textbf{Output:} $\mathbf{C} \in \mathbb{R}^{n \times d}$ distributed in $p$ processes.
\begin{algorithmic}[1]
\Procedure{TS-SpGEMM-Naive}{$\mat{A}_i$, $\mat{B}_i$}  at $P_i$
    \State nzc($\mA_i$) $\gets$ Non-zero column ids in $\mA_i$
    \State $r \gets $ \Call{Alltoall}{nzc($\mA_i$)} \Comment{Requested rows of $\mB_i$}
    \State $\mB_{recv} \gets $ \Call{Alltoall}{$\mB_i(r,:)$}
    \State $\mC_i \gets \Call{LocalSpGEMM}{\mA_i, \mB_{recv}}$
    
\Return $\mC$
\EndProcedure    
\end{algorithmic}
\end{algorithm}

To implement Gustavson's algorithm in a distributed setting, we use 1-D row partitioning of all matrices, where  
$\mA_i {\in}  \mathbb{R}^{\frac{n}{p} \times n}$, $\mB_i {\in}  \mathbb{R}^{\frac{n}{p} \times d}$, and $\mC_i {\in}  \mathbb{R}^{\frac{n}{p} \times d}$ are submatrices of $\mA$, $\mB$, and $\mC$ stored by the $i$th process $P_i$.
Table~\ref{tab:notations} shows the list of notations used in the paper. 

Alg.~\ref{alg:ts-spgemm-naive} shows a naive implementation of TS-SpGEMM where $\mA$ and $\mC$ stays stationary while $\mB$ moves.
We call this algorithm \textproc{TS-SpGEMM-Naive}.
In this algorithm, each process sends requests to potentially all other processes to collect the necessary rows of $\mB$ using two AllToAll communication (Line 3 and 4, Alg.~\ref{alg:ts-spgemm-naive}). After the necessary portion of $\mB$ is received, the output is generated by a local SpGEMM.
We use a toy example in Fig.~\ref{fig:dist-gustavson} to explain the process.
Variants of this algorithm are implemented in popular libraries such as 
PETSc~\cite{Petsc2014} and Trillinos~\cite{Heroux2005}. 
We identify several avenues to optimize this algorithm.


\begin{figure}
		\centering
		\includegraphics[width=0.99\linewidth]{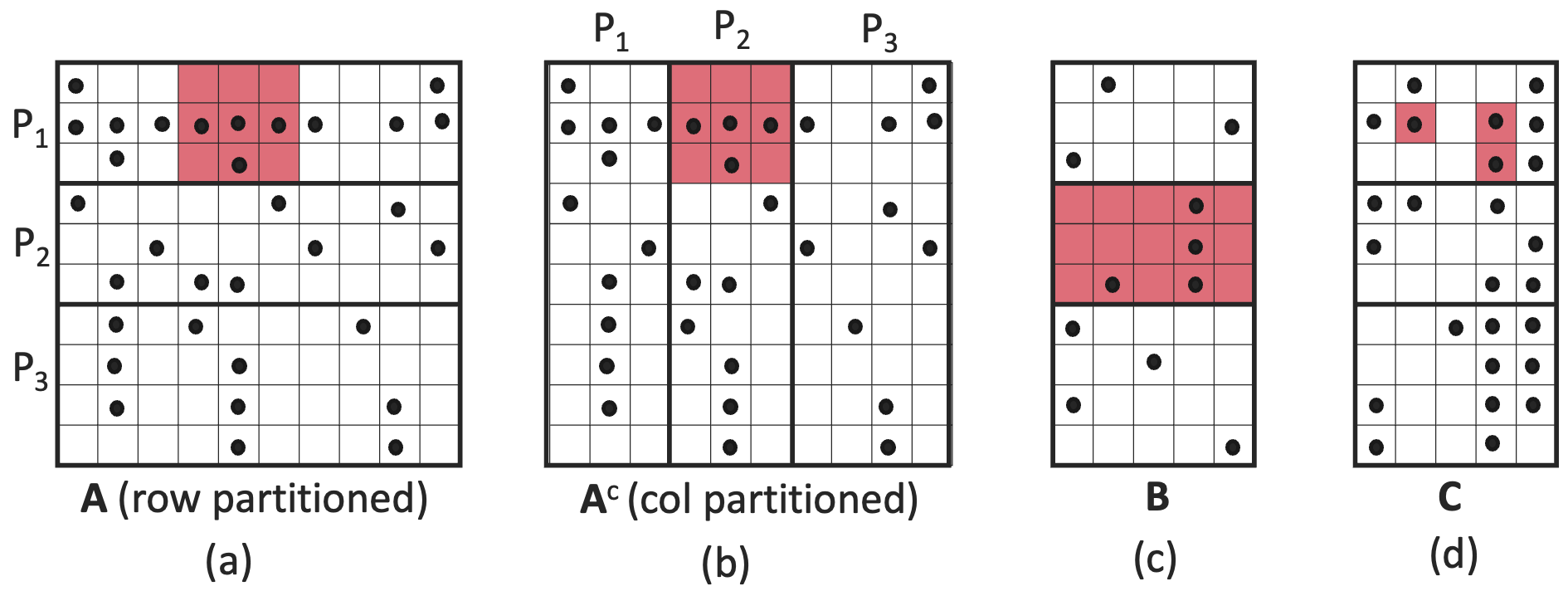}
        \caption{
        Distribution of the same matrices as in Fig.\ref{fig:dist-gustavson} involved in our algorithm. 
        The highlighted regions in each subfigure represent (a) a $3\times3$ tile in $\mA$, (b) the same tile in $\mA^c$, (c) the rows of $\mB$ used by this tile, and (d) the nonzeros in $\mC$ produced by these tiles.
        }
		\label{fig:ts-spgemm-tiling}
\end{figure}

\textbf{Eliminating communication needed to send requests.}
In Alg.~\ref{alg:ts-spgemm-naive} Line 3, we send column indices of $\mA$ to fetch necessary rows of $\mB$.
For example, consider $\mA$ to be an $n\times n$ \erdosrenyi~ matrix of an average degree of $\id{k_A}$.
Then, the expected number of nonzero columns per process would be $\frac{n\id{k_A}}{p}$.
Hence, each process sends $\frac{n\id{k_A}}{p}$ indices in the AllToAll communication at Line 3 of Alg.~\ref{alg:ts-spgemm-naive}.
We eliminate this communication by keeping another copy of $\mA$ and partitioning it column-wise among processes, where $P_i$ stores 
$\mA_{i}^{c}  \in  \mathbb{R}^{ n \times \frac{n}{p}}$.
By utilizing $\mA^c$, each process can precisely identify which rows of its local copy of $\mB$ are needed by other processes, thereby eliminating the necessity to communicate indices beforehand.
Fig.~\ref{fig:ts-spgemm-tiling} explains the benefit of keeping $\mA^{c}$. In this example, $P_2$ is trying to determine which rows of $\mB_2$ will be needed by $P_1$. 
By using $\mA_i^c$, $P_2$ can calculate the necessary rows (in this example all rows of $\mB_2$) that are sent to $P_1$ requiring $P_{1}$ to communicate the corresponding indices.
Thus, our distribution strategy decreases communication at the cost of doubling the memory requirement for $\mA$. 

\textbf{Reducing memory requirement.}
At the end of communication, each process may receive a considerable number of rows of $\mB$ from other processes.
For example, if $\mA_i$ has a dense row, it will need the entire $\mB$ to compute $\mC_i$.
Fig.~\ref{fig:dist-gustavson} illustrates this scenario where the second row of  $\mA_1$ is nearly dense, necessitating all rows except one from $\mB$ to be stored in $P_1$.
This memory bottleneck can also arise when $\mA$ is an \erdosrenyi~ matrix with an average degree of $k_A$.
Let each row of $\mB$ have \id{k_B} nonzeros. 
As the expected number of nonzero columns of $\mA$ in a process is $\frac{n\id{k_A}}{p}$, the total memory requirement in a process could be $\frac{n\id{k_A}\id{k_B}}{p}$.
Thus, depending on the average number of nonzeros in each row of $\mA$ and $\mB$, the memory requirement to receive remote rows of $\mB$ can be prohibitively large.

We address the high-memory requirement, by partitioning $\mA_i$ into tiles.
A $w{\times} h$ tile is a submatrix of $\mA_i$, where  $\mA_i(w,h){\in}\mathbb{R}^{w\times h}$ with $h{\leq} n/p$ and $w {\leq} n$.
We conduct computations tile by tile, where each process receives rows of $\mB$ corresponding to the nonzero columns in the current tile of $\mA$.
Fig.~\ref{fig:ts-spgemm-tiling} shows an example where $P_1$ computes with the highlighted tile of $\mA$ by storing $\mB_2$ (also highlighted in Fig.~\ref{fig:ts-spgemm-tiling}) in $P_1$. We discuss the tile selection policy in the next section.

\textbf{Improving memory locality in computation}
In Line 5 of Alg.~\ref{alg:ts-spgemm-naive}, local matrix multiplication is performed to generate the final output.
Since the local sparse matrices are stored in CSR format, this computation can potentially involve accessing rows of $\mB_{recv}$ randomly depending on the sparsity pattern of $\mA_i$.
By computing tile by tile, we also maintain a reasonable memory locality for this random access pattern.

\textbf{Improving communication by alternating between moving $\mB$ and $\mC$.}
Up to this point, we have exclusively focused on communicating submatrices of $\mB$ needed for a tile for localized computation of $\mC$ within their assigned processes. We refer to this method of computation as the {\em local mode of computation.}
However, if there is a dense row of $\mA_i$ in a tile, 
we can further optimize the communication required to compute the relevant portion of $\mC_i$.
For example, consider the tile highlighted in Fig.~\ref{fig:ts-spgemm-tiling}.
This tile affects the highlighted entries of $\mC$ in $P_1$ (Fig.~\ref{fig:ts-spgemm-tiling}d).
If we perform this computation at $P_1$, $4$ nonzero elements of $\mB$ need to be communicated from $P_2$ to $P_1$ (Fig.~\ref{fig:ts-spgemm-tiling}c), but this computation affects only $3$ non-zeros of $\mC$ (Fig.~\ref{fig:ts-spgemm-tiling}d).
Instead, because we maintain a column partitioned copy $\mA^c$, $P_2$ can compute the relevant entries in $\mC$ and send the result back to $P_1$, reducing the required amount of communication.
 We refer to this method of computation as the {\em remote mode of computation.}
This communication optimization via remote computation only works when the number of output nonzeros produced for a tile is less than the number of nonzeros required from $\mB$.
Due to maintaining two copies of $\mA$, it is possible to mark each tile as a local or remote compute tile through a symbolic step without requiring any communication.
In this way, it is possible to switch between the move-$\mB$ and move-$\mC$ variants of TS-SpGEMM.

\textbf{Improving load balance by alternating between moving $\mB$ and $\mC$.}
Line 5 of Alg.~\ref{alg:ts-spgemm-naive} may cause computation and memory load imbalance when there is a dense row in matrix A. 
Unlike the 2D distribution of matrices, the memory imbalance is inherent in 1D distribution, a characteristic that persists in our algorithm.
However, by delegating computations to remote processes, we can partially balance the computational workload, which enhances the scalability of our algorithm.

\subsection{Distributed TS-SpGEMM with tiling}  \label{sec:spgemm_with_tiling}
In the previous section, we described several modifications of Alg.~\ref{alg:ts-spgemm-naive} to reduce memory and communication requirements and improve load balance. 
This section discusses the improved algorithm with all those improvements.

{\bf Overall data distribution and storage.} 
As mentioned before, we use 1-D row partitioning of all matrices, where  
$\mA_i {\in}  \mathbb{R}^{\frac{n}{p} \times n}$, $\mB_i {\in}  \mathbb{R}^{\frac{n}{p} \times d}$, and $\mC_i {\in}  \mathbb{R}^{\frac{n}{p} \times d}$ are submatrices of $\mA$, $\mB$, and $\mC$ stored the $i$th process $P_i$.
Additionally, we use 1-D column partitioning of $\mA$ where $P_i$ stores $\mA_i^c {\in}  \mathbb{R}^{n \times \frac{n}{p} }$.
At the $i$th process, $\mA_i$ is divided into $w{\times} h$ tiles, where each tile is a submatrix of $\mA_i$  with $h\leq n/p$ and $w \leq n$.
Similarly, $\mA_i^c$ is divided into $h{\times} w$ tiles. 

{\bf Overview of our distributed TS-SpGEMM algorithm.}
Algorithm \ref{alg:distSpGEMM} provides a high-level description of our algorithm from $P_i$'s point of view.
At fist, we generate $w{\times} h$ tiles for $\mA_i$ and $h{\times} w$ tiles for $\mA_i^c$ (Algorithm~\ref{alg:distSpGEMM}, Line~\ref{li:generate_tiles})). The optimal tile width and height depend on the available memory and sparsity of input matrices. 
We tune them empirically as discussed in the result section.
After the tiles are generated, we categorize them in local, remote and diagonal tiles (Algorithm~\ref{alg:distSpGEMM}, Line~\ref{li:tile_modes})). 
The algorithm used to group tiles is discussed in the next section.

\begin{algorithm}[!t]
\caption{Distributed TS-SpGEMM algorithm at $P_i$}
\label{alg:distSpGEMM}
\begin{algorithmic}[1]
\State \textbf{Inputs:}
 \State $\vect{A}_{i}$: Row partition of $\vect{A}$ stored at $P_i$
 \State $\vect{A}_{i}^{c}$: Column partition of $\vect{A}$ stored at $P_i$
 \State $\mB_{i}$: Row partition of $\vect{B}$ stored at $P_i$
 \State $\mathbb{X}$: Tile mode (local/remote) selection  policy
\State \textbf{Output:}
  \State $\mC_{i}$: Row partition of the output stored at $P_i$
\Procedure{Dist-TS-SpGEMM}{$\mA_{i},\mA_{i}^c, \mB_{i},h,w$} at $P_i$
\State $\Call{generateTiles}{\vect{A}_i,\vect{A}_{i}^{c},h,w}$ \label{li:generate_tiles}
\State $\Call{decideMode}{\mA_i,\mA_i^c,\mB_i,\mathbb{X}}$ \label{li:tile_modes}

\LineComment{Compute output for remote  tiles}
\For {each remote tile $\mA^{\text{remote}}_i$ at $P_i$}
   \State $P_j \leftarrow $ the remote computation process for this tile
   \State At $P_j$, extract $\mA^{\text{remote}}_i$ from $\mA^{c}_j$
   \State $ \vect{B}^{\text{remote}}_j \leftarrow$ At $P_j$, extract the submatrix from $\vect{B}_j$ corresponding to nonzero columns of the remote tile \label{li:fetch_remote_data}
  
   \State $ \vect{C}^{\text{remote}}_j \leftarrow \Call{LocalSpGEMM}{\mA^{\text{remote}}_i, \vect{B}^{\text{remote}}_j}$ 
    \State Send $ \vect{C}^{\text{remote}}_j$ back to $P_i$
   \State $\vect{C}_i =  \Call{Merge}{\vect{C}_i,\vect{C}^{\text{remote}}_j}$ at $P_i$ \label{li:comput_local_tile_out}
 \EndFor
 \\
 \LineComment{Compute output for diagonal  tiles}
 \State $ \vect{C}^{\text{diag}}_i \leftarrow \Call{LocalSpGEMM}{\mA^{\text{diag}}_i, \vect{B}_i}$ 
   \State $\vect{C}_i =  \Call{Merge}{\vect{C}_i,\vect{C}^{\text{diag}}_i}$ 
 \\
 \LineComment{Compute output for local  tiles}
\For {each local tile $\mA^{\text{local}}_i$ at $P_i$}
   \State $P_j \leftarrow $ Process storing the necessary $\mB$ submatrix
   \State $ \vect{B}^{\text{local}}_j \leftarrow$ Fetch $\mB$ submatrix from $P_j$ \label{li:fetch_remote_data}
   \State $ \vect{C}^{\text{local}}_i \leftarrow \Call{LocalSpGEMM}{\mA^{\text{local}}_i, \vect{B}^{\text{local}}_j}$ 
   \State $\vect{C}_i =  \Call{Merge}{\vect{C}_i,\vect{C}^{\text{local}}_i}$ \label{li:comput_local_tile_out}
 \EndFor
\label{li:merge_output}
\EndProcedure    
\end{algorithmic}
\end{algorithm}

{\bf Processing remote tiles (lines 11-18 in Algorithm~\ref{alg:distSpGEMM}).}
Let $A_i^{\text{remote}}$ be a $w\times h$ tile stored
at $P_i$.
Let $P_j$ denote the process where the required portion of $\mB$ necessary for this tile is stored\footnote{When the tile width is greater than $n/p$, the necessary rows of $\mB$ is distributed across multiple processes. For simplicity, we refer to a single process $P_j$ in line 13 of Algorithm~\ref{alg:distSpGEMM}}.
Given the column partitioned matrix $\mA^c$, $P_j$ can access the remote tile $A_i^{\text{remote}}$ from $\mA^c$ without any communication with $P_i$.
Then, $P_j$ extracts rows of $\mB_j$ corresponding to nonzero columns of this remote tile and store it in $\mB_j^{\text{remote}}$ (line 15).
After multiplying the remote tile with $\mB_j^{\text{remote}}$, the result is sent back to $P_i$, where the partial result is merged with $\mC_i$. 

{\bf Processing diagonal tiles (lines 20-22 in Algorithm~\ref{alg:distSpGEMM}).}
Let $A_i^{\text{diag}} = \mA_i \cap \mA_i^c$ be a tile on the diagonal of $\mA$.
For this tile, the corresponding entries of $\mB$ are also available in $P_i$. Hence, this multiplication is performed locally without any communication.

{\bf Processing local tiles (lines 24-29 in Algorithm~\ref{alg:distSpGEMM}).}
Let $A_i^{\text{local}}$ be a $w\times h$ tile stored
at $P_i$.
Let $P_j$ denote the process where the required portion of $\mB$ necessary for this tile is stored.
Given the column partitioned matrix $\mA^c$, $P_j$ can access the local tile $A_i^{\text{local}}$ from $\mA^c$ and extracts necessary portion of $\mB_j$ at $\mB_j^{\text{local}}$ (line 27).
Then, $P_j$ send  $\mB_j^{\text{local}}$ to $P_i$.
Upon receiving $\mB_j^{\text{local}}$, $P_i$ performs local computations and merge the results with $\mC_i$.

{\bf Consolidated communication.}
For simplicity, Alg.~\ref{alg:distSpGEMM} discusses communication concerning $P_i$.
As all processes engage in computations for both local and remote tiles, 
communication for the $i$th tile across all processes is consolidated into a single AllToAll communication at lines 17 and 27 of Alg.~\ref{alg:distSpGEMM}.

\subsection{Local computations}
Algorithm~\ref{alg:distSpGEMM} performs two computations: (1) SpGEMM involving local or remote tiles and (2) merge results from a tile with the results from other tiles. 
For both of these operations, we use SPA or hash-based accumulators.
Previous work~\cite{nagasaka2019performance, hussain2022parallel} demonstrated that hash-based merging performs the best for  $\mA\mA$ and $\mA\mA\transpose$ operations. 
However, for tall-and-skinny $\mB$ matrices, we observed that the SPA outperforms hash-based SpGEMM and merging techniques.
This is attributed to the fact that each row of the output matrix is of length $d$. Therefore, a row-by-row SpGEMM necessitates a dense vector for SPA with a length of $d$. When $d$ is small, it can easily fit into the lowest level of cache, resulting in enhanced performance.
We parallelize local computations by assigning different rows of the output to $t$ threads.  
For $d>1024$, we opt for a hash-based SpGEMM, as at large values of $d$, SPA tends to spill out of the cache. 

\begin{figure}[!t]
		\centering
		\includegraphics[width=0.99\linewidth]{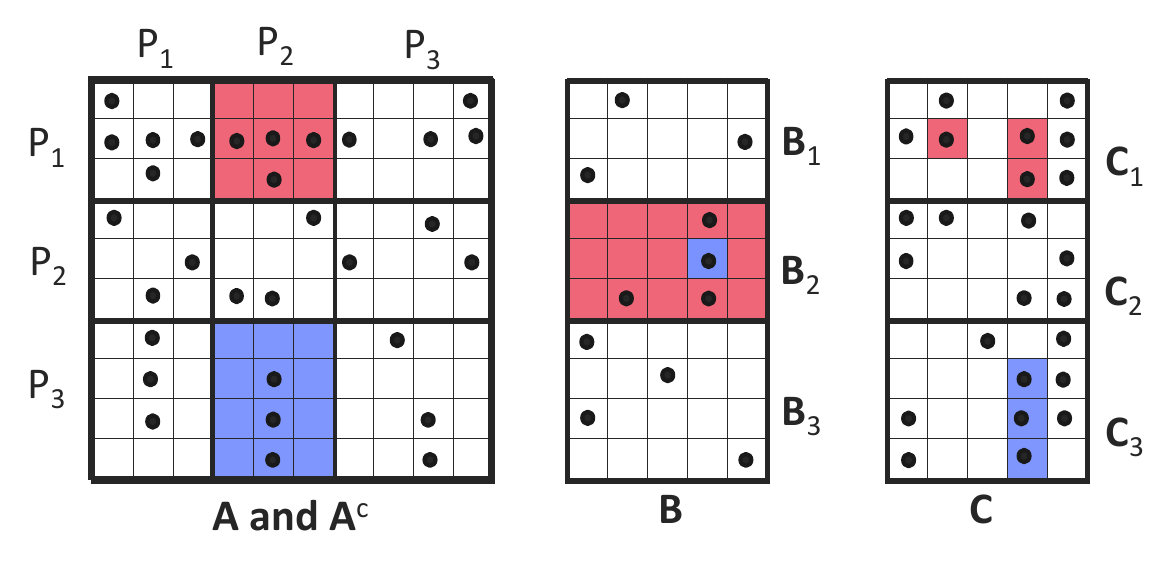}
        \caption{ $P_2$ decides the mode of two tiles shown in red and blue within $\mA$. To compute $\mC_1$, the red tile is multiplied with the entire $\mB_2$.
        Since $\nnz(\mB_2)$ is greater than the affected nonzeros in $\mC_1$, the red tile is marked as a remote tile within $\mA_1$. 
        To compute $\mC_3$, the blue tile is multiplied with just one nonzero in $\mB_2$ shown in blue.
        In this case, it is beneficial to communicate necessary data from $\mB$ and hence, the blue tile is marked as a local tile within $\mA_3$.}
		\label{fig:local-remote}
\end{figure}
\subsection {Tile mode selection}
We use a symbolic step to categorize tiles into local and remote modes with an aim of reducing communication.
In this step, $P_i$ computes the modes of tiles available in $\mA^c_i$ based on the exact communication overheads of possible local and remote computations for each tile.
Fig.~\ref{fig:local-remote} explains this step with three processes.
In this example, the red tile is located in $P_1$ as part of $\mA_1$, but it is also available in $P_2$ as part of $\mA^c_2$.
Hence, $P_2$ can multiply the red tile with its local $\mB_2$ to estimate the contribution of this tile to $\mC_1$.
In Fig.~\ref{fig:local-remote}, the entire $\mB_2$ (four nonzeros) is needed for the red tile to generate three nonzeros in $\mC_1$.
Hence, a remote computation of this red tile at $P_2$ and sending the results back to $P_1$ reduces communication. Hence, the red tile is marked as a remote tile by $P_2$.

On the other hand, the blue tile is located in $P_3$ as part of $\mA_1$, but it is also available in $P_2$ as part of $\mA^c_3$.
Hence, $P_2$ can multiply the red tile with its local $\mB_2$ to estimate the contribution of this tile to $\mC_3$.
In Fig.~\ref{fig:local-remote}, only one nonzero from $\mB_2$ is needed for the red tile to generate three nonzeros in $\mC_3$.
Hence, a local computation of this blue tile at $P_3$ by getting necessary data from $P_2$ reduces communication. Thus the blue tile is marked as a local tile by $P_2$.
Following these steps, every process can categorize tiles from their respective column partition of $\mA^c$.
Note that the selection tiles do not require any communication.
After the modes of all tiles are finalized, the modes of the tiles are shared with all processes via an AllToAll communication step.
The cost of this communication is not significant since it only communicates a binary value (local or remote) for each tile. 

\subsection{Communication and space complexity}
\textbf{Communication complexity of Algorithm \ref{alg:distSpGEMM}.}
The communication complexity depends on the cost of (a) receiving remotely computed outputs (\ref{alg:distSpGEMM}, lines 17), and (b) sending rows from $\mB$ to other processes for local computations (\ref{alg:distSpGEMM}, line 27). 
In both cases, we use AlltoAll collective communication to transfer data.
To analyze the communication complexity, we used the $\alpha-\beta$ model~\cite{thakur2005optimization},
where $\alpha$ is the latency constant corresponding to the fixed cost of communicating a message, and $\beta$ is the inverse bandwidth corresponding to the cost of transmitting one word of data. Consequently, communicating a message of $n$ words takes $\alpha + \beta n$ time. 

Let \id{k_A}, \id{k_B}, and \id{k_C} denote the average number of nonzeros in each row of $\mA$, $\mB$, and $\mC$, respectively.
We consider an $n/p\times n/p$ tile $\mA^{\text{tile}}$ that is multiplied with $\mB^{\text{tile}}$ to generate partial result $\mC^{\text{tile}}$.
Here, $\nnz(\mB^{\text{tile}})=n\id{k_B}/p$ and $\nnz(\mC^{\text{tile}})=n\id{k_C}/p$.
Assuming the pairwise exchange algorithm typical for long messages in MPI implementations, 
the communication cost for a remote tile is $\mathcal{O}{(\alpha p + \beta\frac{(p-1) n \id{k_C} }{p}})$.
Similarly, the communication cost for a local tile is $\mathcal{O}{(\alpha p + \beta\frac{(p-1) n \id{k_B} }{p}})$.
Since a tile is either local or remote depending on the communication cost, the overall communication costs for a tile is $\mathcal{O}{(\alpha p + \beta\frac{(p-1) n \min\{\id{k_B}, \id{k_C}\} }{p}})$.

\textbf{Space complexity of Algorithm \ref{alg:distSpGEMM}.}
In 1-D partitioning, there could be storage imbalance if different row partitions have different numbers of nonzeros. 
This inherent problem of 1-D partitions also exists in our algorithm. 
We analyze the additional memory requirements for an $n/p\times n/p$ tile, which can be translated to other tile sizes. 
We use the nonzero settings discussed in the communication analysis.
For a local tile, the additional memory required to store received submatrices of $\mB$ is $\mathcal{O}(n\id{k_B}/p)$.
Similarly, for a remote tile, the additional memory required to store partial results of $\mC$ is $\mathcal{O}(n\id{k_C}/p)$.
These estimates are not precise, as the actual memory requirements can be substantially lower depending on the sparsity of the tile.
For local SpGEMM, the memory requirement depends on the accumulator (SPA/Hash) used.
For SPA, where each of the $t$ threads maintain their private SPA, the memory requirement is $\mathcal{O}{\left(t\times d + \frac{nnz(\vect{C})}{p}+n\right)}$, and for hash-based SpGEMM the memory complexity is $\mathcal{O}{\left( \frac{nnz(\vect{C})}{p}+n\right)}$. 

\section{Algorithms implemented with TS-SpGEMM}
To demonstrate the utility of TS-SpGEMM in practical settings, we implemented two graph algorithms that require the multiplication of a square matrix with a tall-and-skinny matrix. 
We briefly discuss these algorithms in this section.

\subsection{Distributed multi-source BFS}
\begin{algorithm}[!t]
\caption{Multi-source BFS}
\label{alg:msbfs}
\textbf{Input:} Adjacency matrix $\mathbf{A} \in \mathbb{B}^{n \times n}$ and vector $\mathbf{f} \in \mathbb{B}^{n \times 1}$ with d non-zero entries representing sources of BFS traversal.
\textbf{Output:} Tall and Skinny matrix $\mathbf{S} \in \mathbb{B}^{n \times d}$ representing vertices reachable from d sources.
\begin{algorithmic}[1]
\Procedure{Dist-MSBFS}{$\mat{A}$, $\mat{f}$}
\State $ \mathbf{F} \gets$ \Call{Init}{$\mat{f}$} \Comment{Initialize frontier, $\mathbf{F} \in \mathbb{B}^{n \times d}$}
\State $\mathbf{S} \gets \mathbf{F}$ \Comment{Mark sources as visited}
\State SR $\gets$ \Call{Semiring}{$\land, \lor$}
\While{\textit{nnz}($\mathbf{F}$) $> 0$ }
    \State $\mathbf{N} \gets$ \Call{TS-SpGEMM}{$\mathbf{A}, \mathbf{F}$, SR} \Comment{Discover next frontier}
    \State $\mathbf{F} \gets \mathbf{N} \setminus \mathbf{S}$ \Comment{Remove already visited vertices}
    \State $\mathbf{S} \gets \mathbf{S} \lor \mathbf{N} $ \Comment{Update so far visited list}
\EndWhile
\State \Return $\mathbf{S}$
\EndProcedure    
\end{algorithmic}
\end{algorithm}
BFS traversal from a single source can be translated as a sequence of sparse matrix-sparse vector (SpMSpV) operations~\cite{spmspvipdps17} with ($\land, \lor$) semiring (or a (sel2nd,min) semiring when the reconstruction of the BFS tree is desired)~\cite{mathgraphblas16, davis2019algorithm}.
In this formulation, the input vector represents the current BFS frontier, while the output vector, after eliminating already visited vertices, indicates the next frontier.
In the beginning, the input vector contains exactly one non-zero, indicating the source of the traversal.
For scale-free graphs, the BFS frontier initially becomes denser and then gradually becomes sparser as more vertices are discovered~\cite{beamer2012direction}.

\begin{figure*}[!t]
		\centering
		\includegraphics[width=0.99\textwidth]{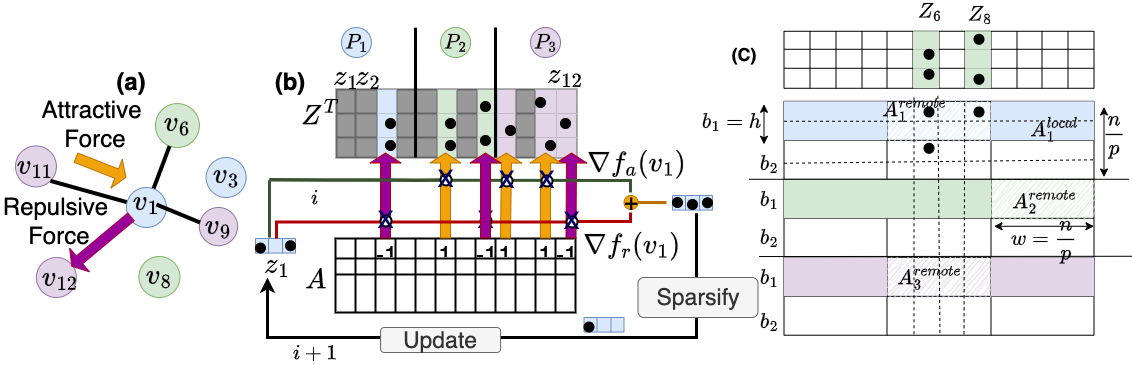}
		\caption{(a) An illustration of the force-directed node embedding, where neighboring vertices generate attractive forces and non-neighboring vertices generate repulsive forces. (b) The matrix $\mA$ represents the adjacency matrix of the graph and $\vect{Z}\transpose$ represents the sparse embedding matrix. Then, force computations are mapped to a TS-SpGEMM operation. (c) The minibatch computation, where we set tile height to be equal to the batch size.}
		\label{fig:dist_sparse_embedding}
\end{figure*}

A multi-source BFS from $d$ sources runs concurrent BFSs on the same input graph as described in Alg.~\ref{alg:msbfs}.
When implemented with TS-SpGEMM, the multi-source BFS maintains a tall-and-skinny sparse matrix $\mathbf{F}$ where the $i$th column represents the frontier corresponding to the $i$th source vertex.
To keep track of visited vertices, we maintain a tall-and-skinny sparse matrix $\mathbf{S}$ where the $i$th column represents all vertices visited from the $i$th source vertex. 
At first, we build $\mathbf{F}$ from the set of source vertices such that each column of $\mathbf{F}$ has just one nonzero (Line 2, Alg.~\ref{alg:msbfs}).
We continue expanding frontiers until $\mathbf{F}$ is empty.
In each iteration, we discover the vertices reachable from the current frontier through the multiplication of the adjacency matrix by the frontier matrix with ($\land, \lor$) semiring (Line 6, Alg.~\ref{alg:msbfs}).
After each multiplication, we remove already visited vertices from $\mathbf{F}$ and then update the set of visited vertices across all BFSs (Line 7 and 8, Alg.~\ref{alg:msbfs})
The updated frontier $\mathbf{F}$ becomes the input to the next iteration.
As the sparsity of matrix $\mathbf{F}$ changes significantly throughout iterations, this algorithm serves as an excellent testing ground for TS-SpGEMM.

\subsection{Distributed sparse embedding}
\label{sec:sparse-emebdding}
We consider a node embedding problem where each vertex in a graph is embedded in a $d$-dimensional vector space.
Typically, the embedding matrix $\vect{Z} \in \mathbb{R}^{n \times d}$ is a dense matrix, but it can be sparsified without compromising the quality of embedding.
Hence, in sparse embedding, the embedding matrix $\vect{Z} \in \mathbb{R}^{n \times d}$ takes the form of a tall-and-skinny sparse matrix, presenting an interesting application for our TS-SpGEMM algorithm.
 
We aim to implement a force-directed node embedding algorithm called Force2Vec~\cite{Force2Vec} that uses attractive and repulsive forces among vertices to compute embeddings. 
We specifically choose Force2Vec to demonstrate TS-SpGEMM due to its existing implementation with SpMM~\cite{ranawaka2024scalable}. By inducing sparsity in the embedding matrix, we develop a sparse variant of Force2Vec using distributed TS-SpGEMM.
Figure \ref{fig:dist_sparse_embedding}  shows a sample graph (left figure) and matrix representation with embedding computation (right figure). We use synchronous SGD with negative samples to compute the embedding. The matrix $\vect{A} \in \mathbb{R}^{n \times n}$ represents the adjacency matrix of the graph ${G(V,E)}$, where $1$ indicates the neighbor vertices and $-1$ indicates the negative sampled non-neighbor vertices. The embedding matrices $\vect{Z} \in \mathbb{R}^{n \times d}$, and $\vect{Z^{T}} \in \mathbb{R}^{d \times n}$ represents the tall-and-skinny embedding matrix and it's transpose, respectively. Both $\vect{A}$, $\vect{Z}$, and $\vect{Z^{T}}$ are 1-D partitioned and stored in each process in CSR format. 

As depicted in Figure \ref{fig:dist_sparse_embedding}, each embedding vector in a minibatch is updated parallelly by calculating the attractive force gradient $\Delta f_{a}(v_{i})$, and repulsive force gradient $\Delta f_{r}(v_{i})$ through a sequence of SpGEMM operations of \algoname and update each embedding vector using SGD. Afterward, the updated embedding matrix is sparsified by selecting the required number of nonzero entries to achieve the target sparsity by keeping the highest valued entries. This sparsified output is used as the input to the next iteration.

 We set the batch size to the height of a tile, that enables minibatch SpGEMM for \algoname. Reducing the height of a tile increases the communication volume. For instance, in the sub-figure (c) of Figure \ref{fig:dist_sparse_embedding}, the embedding vector $Z_{6}$ needs to be fetched twice in batches $b_{1}$, and $b_{2}$. But, if  $A^{remote}_{1}$ is computed remotely on $P_{2}$, the $Z_{6}$ only needs to be fetched once while computing $b_{2}$. Hence, the remote computations can reduce communication overhead that is incurred in the minibatch scenarios with tiling.

%% file: results.tex
\section{Results}
\begin{table}[!t]
\centering
\caption{Default parameters used in our experiments.}
\label{table:parameters}
\begin{tabular}{l l l l l} 
\toprule
\textbf{Parameter}                 & \textbf{Value}    \\               
\toprule
Number of OpenMP    threads per process                          &16        \\          

Number of processes per node                          &8   \\

Number of processes for application testing        &64 \\

Dimension of $\vect{B}$ matrix $(d)$                             &128 \\

Height of a tile  ($h$)             &$\frac{n}{p}$\\
Width of a tile ($w$)            & $16 \times \frac{n}{p}$\\

Default sparsity of $\vect{B}$                   &80\% \\
Embedding mini-batch size  ($b$)                              & 256  \\

Embedding learning  rate               &0.02 \\

\bottomrule
\end{tabular}

\label{tab:parameters}
\end{table}

\subsection{Experimental setup}
Table \ref{tab:parameters} shows the default parameters used in the experiments. 
We identified these default parameters via extensive benchmarking. 
Users can use default parameters to obtain good performance. 
In particular, we observed that a tile width of ($16\times n/p$) and a tile height of $n/p$ perform the best for most matrices.
We use these default parameters in all experiments unless otherwise stated.
Runtimes were reported as the average of five runs.

\textbf{Experimental platforms.}
We evaluate the performance of our algorithm on the CPU partitions of the NERSC Perlmutter supercomputer. 
A single compute node of the Perlmutter CPU partition is equipped with two AMD EPYC 7763 CPUs with 64 cores and 512GB of memory. 
To compile the programs, SUSE Linux g++ compiler 12.3.0 is used with -O3 option.
We used MPI+OpenMP hybrid parallelization for all experiments.
For in-node multithreading, we experimented with various settings and found that 16 threads per process gave the best performance
Unless otherwise stated, we used $8$ MPI processes per node and $16$ OpenMP threads per process.
For MPI implementation, we used Cray-MPICH-8.1.28.

\begin{table}[!t]
\centering
\caption{Datasets used in our experiments.}
\label{table:dataset_table}
\begin{tabular}{l l l  l l } 
\toprule
\textbf{Dataset} &\textbf{Alias} & \textbf{\# Vertices} & \textbf{\# Edges}    & \textbf{Avg Degree} \\ 
\toprule
pubMed   &pubmed &19,717 &44,338   &4.49\\

flicker &flicker  &89,250 & 899,756 &20.16\\

cora &cora  &2708  &5429  &2 \\

citeseer & citeseer &3312   &4732 &1.4\\

arabic-2005 &arabic     &22,744,080         &639,999,458	 &28.1 \\

it-2004  &it    &41,291,594&1,150,725,436	   &27.8 \\ 

GAP-web  &gap    &50,636,151  &1,930,292,948	  &38.1 \\ 

uk-2002   &uk   &18,520,486	  &298,113,762	    &16.0 \\ 
\erdosrenyi~    &ER   &40000000	  &320000000	    &8 \\
\bottomrule
\end{tabular}
\end{table} 

{\bf Datasets.}
Table.\ref{table:dataset_table} describes the graphs used in our experiments.
We collected these graphs from SNAP~\cite{leskovec2016snap} and the Suitesparse Matrix Colelction~\cite{davis2011university}. 
Additionally, we generate uniformly random tall-and-skinny matrices for our experiments.
In our experiments, the tall-and-skinny matrix  $\mB$ with $s\%$ sparsity means $s\%$ entries in each row of the $\mB$ are zero.

\textbf{Baselines.}
We compare our implementation with state-of-the-art algorithms such  as 2-D Sparse SUMMA~\cite{Buluc2012}, 3-D Sparse SUMMA~\cite{3dspgemmsisc16}, and 1-D  algorithm in PETSc~\cite{Petsc2014}. 
We use 2-D and 3-D implementation available in CombBLAS-2.0\cite{combblas2} and 1-D implementation available in PETSc-3.19.3.


\begin{figure}[!t]
		\centering
		\includegraphics[width=0.99\linewidth]{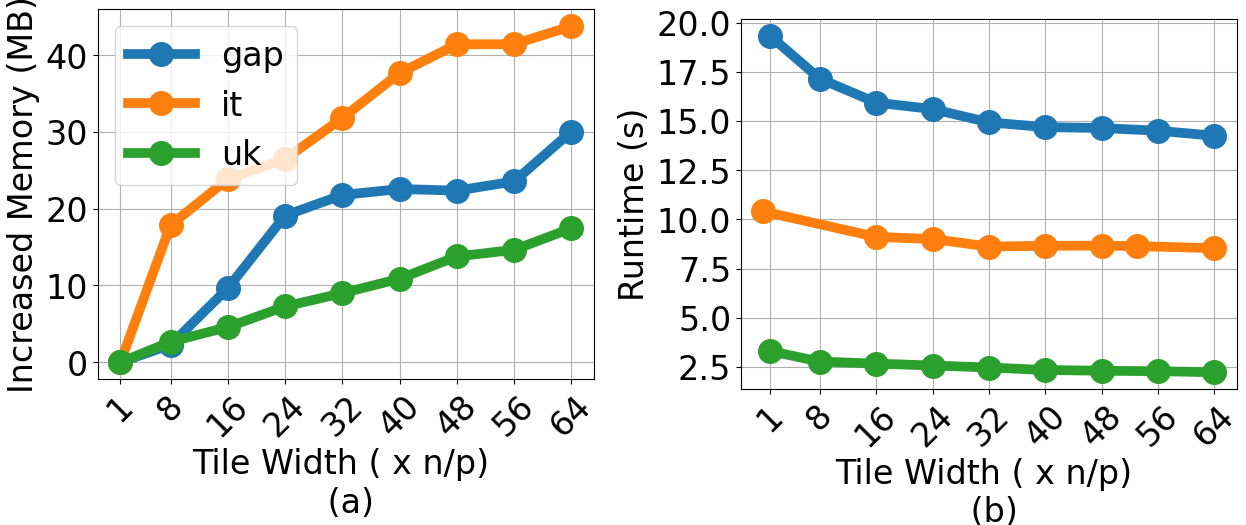}
		\caption{The impact of an increasing tile width on memory and runtime on $8$ nodes ($64$ processes). The x-axis shows the width of the tile ranging from $n/p$ to $n$, expressed as multiples of $n/p$.
        The left subfigure shows the increase in memory consumption, while the right subfigure shows the impact on the runtime.}
	\label{fig:tiling_benchmark}
\end{figure}
\begin{figure}[!t]
		\centering
		\includegraphics[scale=0.3]{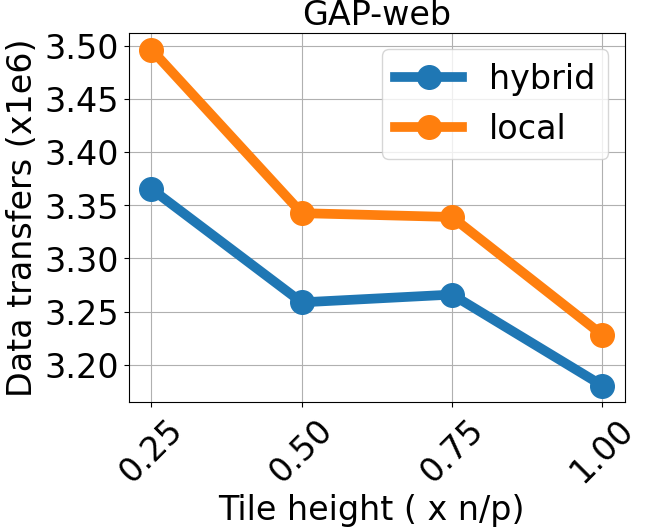}
		\caption{The reduction in data transfers for hybrid mode and local mode. The hybrid mode enables both local and remote tiles, whereas the local mode only uses local tiles. We ran the experiments on $8$ nodes for GAP-web.}
		\label{fig:df_saving_with_height}
\end{figure}
\subsection{Impacts of the tile width}
To determine the optimal tile size, we ran experiments on $8$ nodes ($64$ MPI processes) with three datasets. 
We use the maximum height  $h=n/p$ and vary the width of a tile $w$ from $n/p$ to $n$. The results are shown in Figure~\ref{fig:tiling_benchmark}, where tile widths are shown as multiples of $n/p$. Figure~\ref{fig:tiling_benchmark}(a) shows that the memory consumption increases monotonically with the increase of tile width. This behavior is expected since a higher value of $w$ requires fetching a larger fraction of $\vect{B}$ into the local process.
Hence, to reduce memory consumption, we should use smaller tile widths. However, as shown in Figure~\ref{fig:tiling_benchmark}(b), a small tile width results in longer runtimes due to increased communication rounds. 
Based on the observations from Fig.~\ref{fig:tiling_benchmark}, we determine that $16{\times}n/p$ represents an optimal tile width for achieving the fastest performance with a manageable memory overhead. Hence, we use $16{\times}n/p$ as the default tile width for our experiments.
Furthermore, we tested the data transfer cost for different tile heights, by fixing the tile width to $16{\times}n/p$. Figure \ref{fig:df_saving_with_height} shows that in the hybrid mode where both remote and local computations are enabled, TS-SpGEMM 
reduces the data transfer cost compared to the pure local mode. A small tile height is useful to capture the computations in the sparse embedding application.
 
\subsection{SpGEMM vs SpMM}
\begin{figure}[!t]
		\centering
		\includegraphics[width=0.99\linewidth]{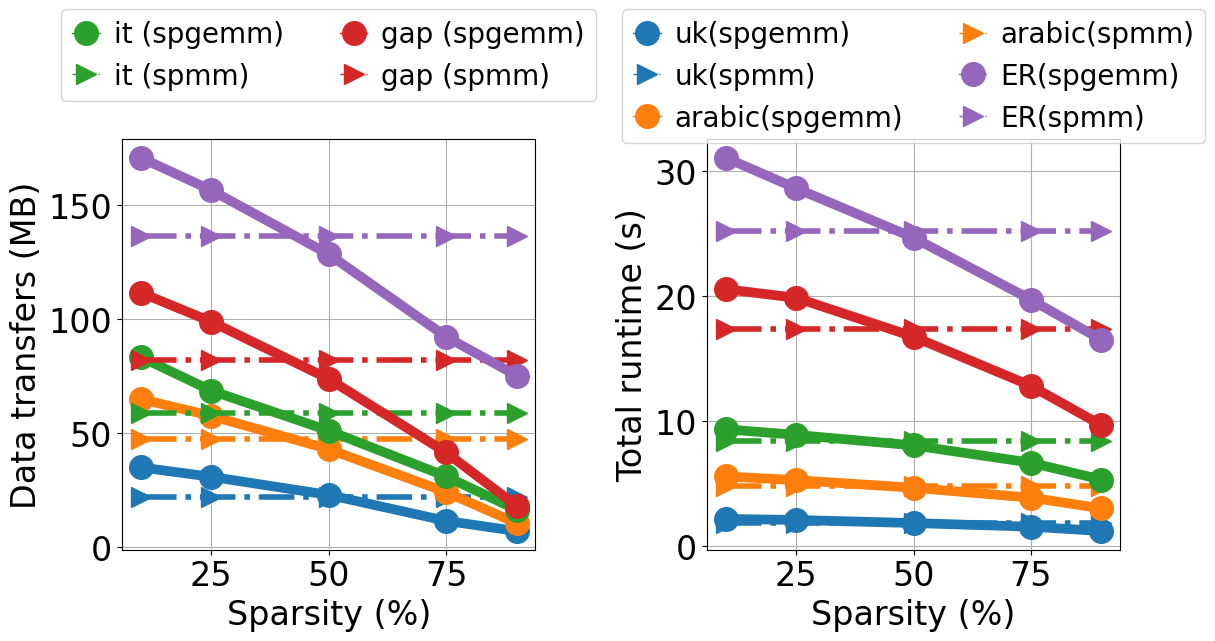}
		\caption{The Figure shows the comparison results of the \algoname and SpMM version. The left sub-figure shows the communication volume and the right sub-figure shows the runtime for different sparsity levels.}
		\label{fig:spmm_spgemm_comp}
\end{figure}

\begin{figure*}[!tb]
		\centering
		\includegraphics[width=0.99\textwidth]{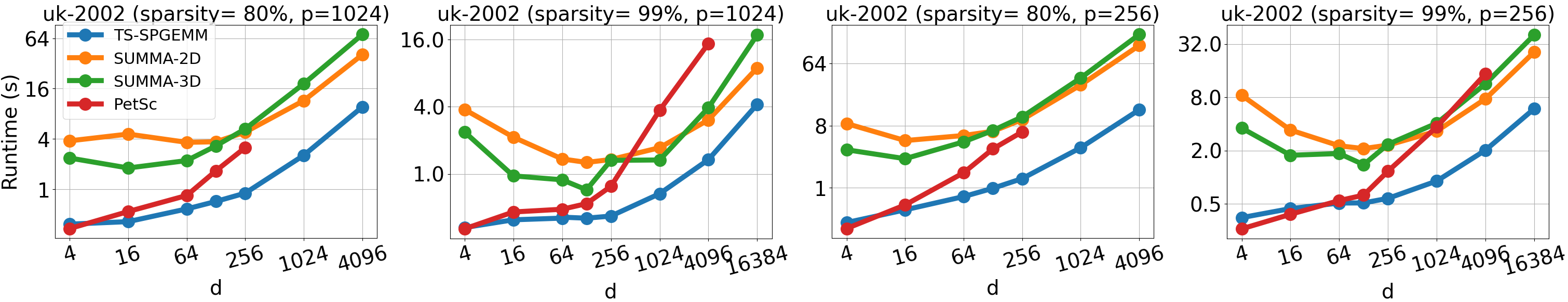}
		\caption{This figure shows the runtime comparison of the \algoname, 2-D SUMMA, 3-D SUMMA, and  PETSc (1-D) for \textit{uk-2002} dataset under different dimensions for sparsity $80\%$, and $99\%$ on $128$ nodes ($p=1024$) and $32$ nodes ($p=256$). For 80\% sparsity, we encountered out-of-memory issues with the PETSc binary converter, preventing us from converting inputs to binary format for dimensions $d\geq256$.}
		\label{fig:varying_d_exp}
\end{figure*}
\begin{figure*}[!t]
		\centering
		\includegraphics[width=0.99\textwidth]{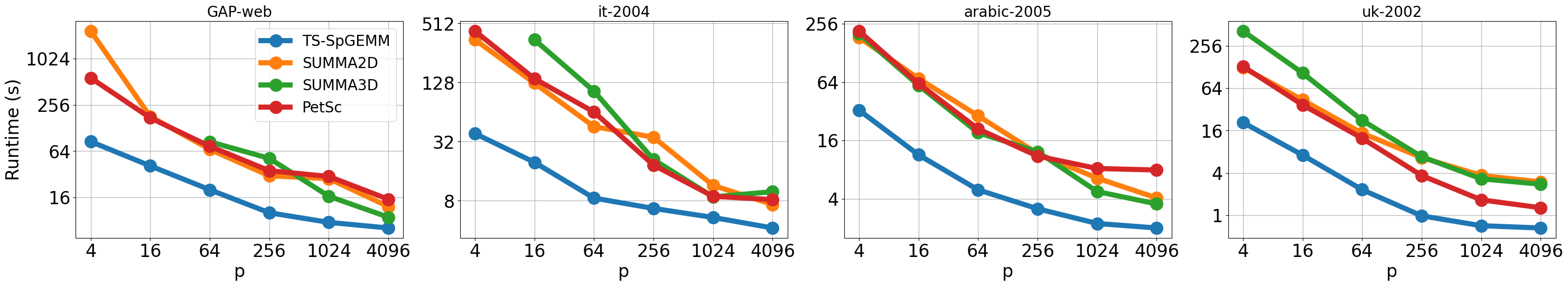}
		\caption{The figures show the strong scaling runtime for \algoname 2-D SUMMA, 3-D SUMMA, and PetSc for \textit{GAP-web}, \textit{it-2004}, \textit{arabic-2005} and \textit{uk-2002} datasets. We ran this experiment for $\vect{B}$ with 128 dimensions and $80\%$ sparsity.}
		\label{fig:scalability_sp_80}
\end{figure*}
\begin{figure*}[!t]
		\centering
		\includegraphics[width=0.99\textwidth]{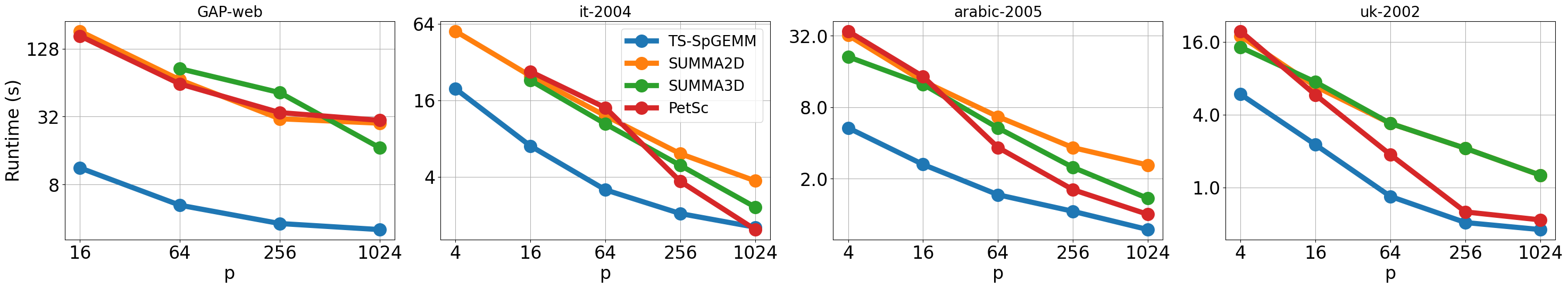}
		\caption{The figures show the strong scaling runtime for \algoname, 2-D SUMMA, 3-D SUMMA, and PetSc for \textit{GAP-web}, \textit{it-2004}, \textit{arabic-2005}, and \textit{uk-2002} datasets. We ran this experiment for $\vect{B}$ matrix with 128 dimensions and $99\%$ sparsity.}
		\label{fig:scalability_sp_99}
\end{figure*}

\begin{figure*}[!t]
		\centering
		\includegraphics[width=0.99\textwidth]{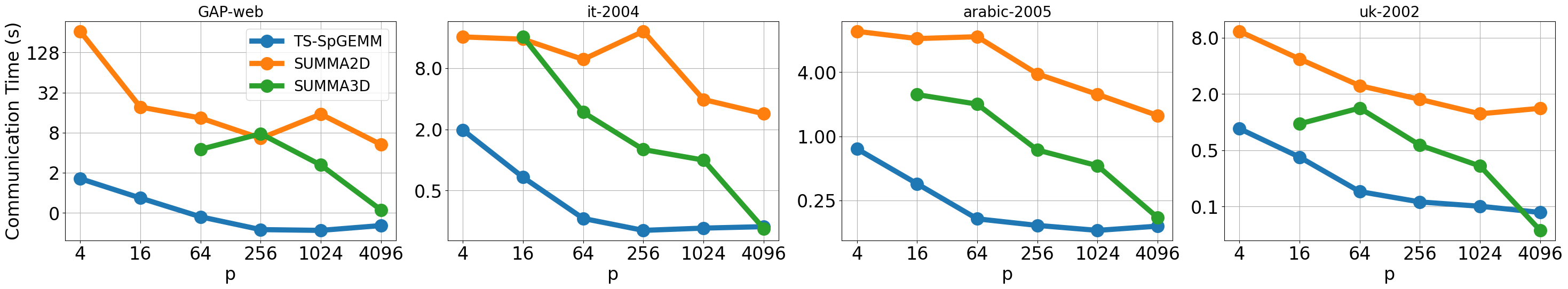}

		\caption{The figures show the strong scaling communication time for \algoname, 2-D SUMMA, 3-D SUMMA, and PetSc for \textit{GAP-web}, \textit{it-2004}, \textit{arabic-2005}, and \textit{uk-2002} datasets. We ran this experiment for $\vect{B}$ matrix with 128 dimensions and $80\%$ sparsity.}
		\label{fig:scalability_sp_com_80}
\end{figure*}

When the tall-and-skinny input matrix ($\mB$) is sufficiently dense, running an SpMM may run faster than TS-SpGEMM. To determine the sparsity threshold of $\vect{B}$ at which TS-SpGEMM begins to outperform SpMM, we implemented an SpMM with a dense $\mB$ using the same communication patterns as TS-SpGEMM. 
We confirmed that our SpMM performs comparably or better than the 1.5D dense shifting algorithm~\cite{selvitopi21_tallskinnydense, block2024two}.
 We ran this experiment on $32$ nodes with $256$ MPI processes and show the results in Figure \ref{fig:spmm_spgemm_comp}. 
For all datasets with sparsity exceeding $50\%$, TS-SpGEMM communicates less data and runs faster than SpMM. 
This sparsity threshold is justified by considering that TS-SpGEMM requires communication of both indices and values, whereas SpMM only communicates values.
The reduction in communication may not translate into a proportional reduction of runtime because local SpGEMM computation is more costly than SpMM.
Furthermore, the total runtime of SpGEMM is higher up to $50\%$ sparsity compared to SpMM due to additional overheads involved with random memory access and computation. We recommend using \algoname only for applications where the second matrix is at least $50\%$ sparse.

\begin{figure*}[!t]
		\centering
		\includegraphics[width=0.99\textwidth]{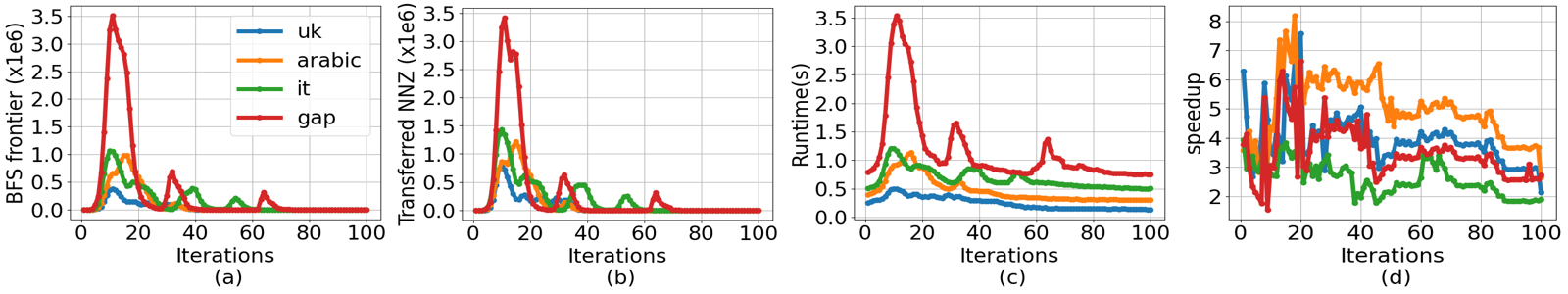}
		\caption{The multi-source BFS implementation using \algoname ran on $8$ nodes (64 MPI processes). The sub-figure (a) shows the average BFS frontier. The sub-figure (b) shows the average communicated nnz. The sub-figure (c) shows the runtime for each iteration. The sub-figure (d) shows the speedup with respect to 2-D SUMMA on each iteration.}
		\label{fig:bfs_results}
\end{figure*}

\begin{figure*}[!t]
		\centering
		\includegraphics[width=0.99\textwidth]{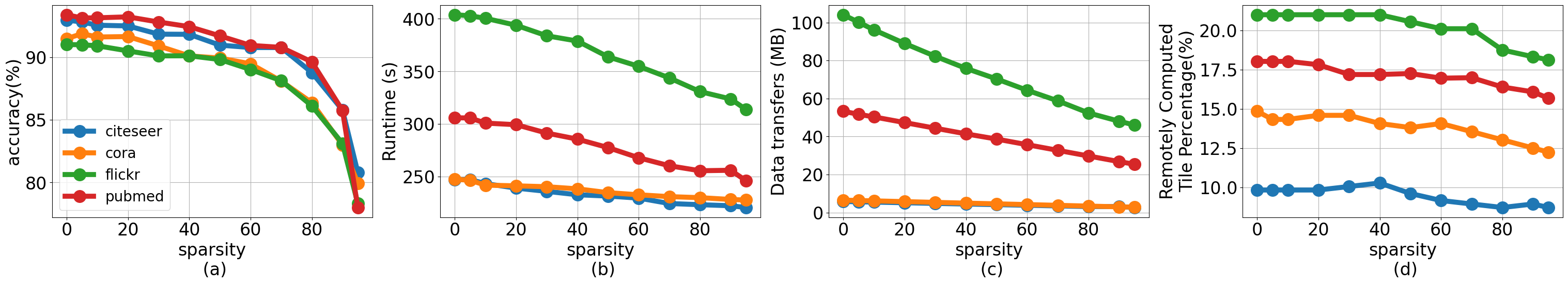}
		\caption{The performance of the sparse embedding algorithm on $8$ nodes (64 MPI processes).  (a) The link-prediction accuracy for  \textit{citeseer}, \textit{cora}, \textit{flicker}, and \textit{pubmed} with varying sparsity in the embedding matrix. (b) Total runtime with varying sparsity. (c) The communicated volume. (d) The percentage of remotely computed tiles.}
		\label{fig:embedding_results}
\end{figure*}

\subsection{Comparison with other distributed SpGEMM}
We compared \algoname with state-of-the-art implementations: PETSc (1-D algorithm), 2-D Sparse SUMMA, and 3-D Sparse SUMMA. We ran the experiments by varying $d$ under $80\%$ and $99\%$ sparsity in $32$ nodes and $128$ nodes. The results are shown in the Figure \ref{fig:varying_d_exp}. Our algorithm outperforms the others when varying the dimension of the second matrix $\vect{B}$ from $4$ to $16,384$. According to the results, PETSc and \algoname have similar runtimes at $d{=}4$, but PETSc SpGEMM performance drops noticeably at around $d{=}64$.  
This is because, at $d{=}4$ with 80\% sparsity, it is feasible to store the entire $\mB$ on a single process. Hence, there is no significant benefit of tiling when $d$ is very small.
By contrast, SUMMA-2D and SUMMA-3D do not perform well when $d<256$, but their runtimes appear to be more competitive at higher $d$. 
The behavior of SUMMA-2D and SUMMA-3D is also predictable because these algorithms involve communication for both $\mA$ and $\mB$. In cases where $d$ is small and $\mB$ is sparse, it is advantageous to communicate only $\mB$, as our algorithm and PETSc do.
A more interesting result is TS-SpGEMM's superior performance at $d=4,096$ and $d=16,384$ where 1D SpGEMM in PETSc performs poorly. 
This improved performance of TS-SpGEMM is attributed to tiling and other optimizations discussed in the method section. 
Thus, Figure \ref{fig:varying_d_exp} demonstrates that \algoname outperforms state-of-the-art algorithms when $\vect{B}$ is a tall-and-skinny matrix.

\subsection{Scalability}
We ran scalability tests ranging from $1$ node to $512$ nodes for sparsity levels of $80\%$ and $99\%$. The results are shown in Figures \ref{fig:scalability_sp_80} and \ref{fig:scalability_sp_99} respectively. \algoname scales up to $512$ nodes (4096 MPI processes; 65,536 cores) and outperforms other SpGEMM implementations.  Runtime scales almost linearly until 1024 processes for both sparsity levels. Past this point, performance scaling has been reduced due to workload reduction. Figure~\ref{fig:scalability_sp_com_80} shows the results for communication scalability for $80\%$ sparsity. TS-SpGEMM's communication scales up to 1024 MPI processes, after which latency begins to dominate.  We did not plot the PETSc communication overhead as it does not report the communication time separately. 
Since SUMMA3D is a communication-avoiding algorithm, its communication scales much better than other algorithms. 
SUMMA3D communication can even beat \algoname at $512$ nodes by utilizing more layers. Still, applications relevant to \algoname do not typically need more than 128 nodes to run, as the second matrix stays as tall and skinny.

\subsection{Multi-source BFS}
The multi-source BFS is the first application we implemented using \algoname. We test the application using $8$ nodes (64 MPI processes). 
We use four datasets to evaluate the BFS: \textit{uk-2002}, \textit{arabic-2005}, \textit{it-2004}, and \textit{GAP-web}. 
We consider $128$ sources randomly selected as the starting nodes. Hence, $\vect{B}$ is of size $n\times 128$ and initially contains one randomly chosen non-zero per column. The evaluation results are given in the Figure \ref{fig:bfs_results}. In the BFS implementation, we sparsify the output of each iteration such that it only contains the newly visited vertices and feed that output as input to the next iteration. 
We can see that the BFS frontier becomes denser only in a few iterations but remains sparse for the rest of the iterations. If there are multiple connected components, it is possible to have several peaks in the BFS frontier for the same dataset. 
If sparsity is greater than $50\%$, we use SpGEMM for the computation, otherwise we can utilize the SpMM version of the TS-SpGEMM. The communication and runtime closely follow the BFS frontier. Figure~\ref{fig:bfs_results}(d) shows the speedup with respect to multi-source BFS implemented with 2-D SUMMA in CombBLAS. We can achieve up to a $10$x speedup in some iterations and around a 5x speedup on average.

\subsection{Sparse Embedding}
As our final application, we implemented a sparse embedding algorithm discussed in Section~\ref{sec:sparse-emebdding}. 
For improved accuracy, we use mini-batch SpGEMM where we set a batch size of $b=0.5\times \frac{n}{p}$. 
The tile height matches the batch size in the minibatch setting.
We ran all the experiments on $8$ nodes (64 MPI processes) and calculated the link prediction accuracy as given by the Force2Vec embedding algorithm~\cite{Force2Vec}. Figure \ref{fig:embedding_results}  shows that we can make the embedding $80\%$ sparse by sacrificing less than 5\% accuracy in link prediction. The runtime and data transfer plots reveal that we can achieve faster convergence and lesser communication overhead with increasing sparsity. Furthermore, sub-figure \ref{fig:embedding_results}(d)  shows that remote tiles play important roles in the minibatch setting.

%% file: conclusion.tex
\section{Conclusion}

Popular distributed algorithms deliver
suboptimal performance for SpGEMM settings where one matrix is 
square and the other is tall and skinny---a variant we call TS-SpGEMM. 
To address this limitation, we 
developed a novel distributed-memory algorithm for TS-SpGEMM that 
employs customized 1D partitioning for all matrices 
and leverages sparsity-aware tiling for efficient data transfers. 
At lower to moderate node counts (up to 128 nodes), TS-SpGEMM shows superior performance compared to 1D, 2D, and 3D SpGEMM algorithms. This trend persists even at 512 nodes, highlighting the effectiveness of our optimizations.
Further, we use our algorithm to
implement multi-source BFS and sparse graph embedding algorithms
and demonstrate their scalability up to 512 Nodes on NERSC Perlmutter.

One limitation of our TS-SPGEMM algorithm is that it requires storing two copies of the first input matrix, $\mA$, which increases the overall memory usage. 
However, most communication-avoiding algorithms, including SUMMA3D, use additional memory to reduce communication overhead. 
We believe that most graph and sparse matrix algorithms can accommodate an extra copy of $\mA$ in a distributed-memory system. 
Another limitation of our algorithm is the use of 1D matrix partitioning, which can lead to load imbalances in scale-free graphs that typically have denser rows. 
While the memory imbalance is inherent to 1D partitioning, we addressed the computational imbalance using virtual 2D partitioning, which performs multiplication tile by tile. Nevertheless, the memory imbalance associated with input matrices can still pose challenges for scale-free graphs.

The optimizations used in TS-SpGEMM can be adapted to distributed SpMM~\cite{selvitopi21_tallskinnydense, block2024two} and fused matrix multiplication~\cite{rahman2021fusedmm} algorithms.  We observed that while TS-SpGEMM is not faster than SpMM when matrix $\mB$ is fully dense, it outperforms SpMM when $\mB$ is 50\% or more sparse. Additionally, TS-SpGEMM is not the optimal choice when $\mB$ closely resembles $\mA$ in shape and sparsity; however, it still outperforms SUMMA when multiplying a sparse matrix by another sparse matrix that is not tall and skinny.

\section{Acknowledgements}
This research was funded in part by DOE grants DE-SC0022098 and DE-SC0023349; by NSF grants PPoSS CCF 2316233 and OAC-2339607; by SRC JUMP 2.0 ACE Center; and by the National Science Foundation Graduate Research Fellowship Program under Grant No. DGE 21-46756.